\documentclass[%
superscriptaddress,
preprint,
 amsmath,amssymb,
aps,
]{revtex4}

\usepackage{graphicx}
\usepackage{dcolumn}
\usepackage{bm}
\usepackage{xcolor}
\usepackage{float}

\begin{document}


\title{Scattering Cancellation Technique for Acoustic Spinning Objects}

\author{M. Farhat}
\email{mohamed.farhat@kaust.edu.sa}
\affiliation{Computer, Electrical, and Mathematical Science and Engineering (CEMSE) Division, King Abdullah University of Science and Technology (KAUST), Thuwal 23955-6900, Saudi Arabia}


\author{S. Guenneau}
\affiliation{UMI 2004 Abraham de Moivre-CNRS, Imperial College London, London SW7 2AZ, United Kingdom.}

\author{A. Al\`u}%
\affiliation{Photonics Initiative, Advanced Science Research Center, City University of New York, New York, New York 10031, USA}

\author{Y. Wu}
\email{ying.wu@kaust.edu.sa}
\affiliation{Computer, Electrical, and Mathematical Science and Engineering (CEMSE) Division, King Abdullah University of Science and Technology (KAUST), Thuwal 23955-6900, Saudi Arabia}

\begin{abstract}
The scattering cancellation technique (SCT) has proved to be an effective way to render static objects invisible to electromagnetic and acoustic waves. However, rotating cylindrical or spherical objects possess additional peculiar scattering features that cannot be cancelled by regular SCT-based cloaks. Here, a generalized SCT theory to cloak spinning objects, and hide them from static observers, based on rotating shells with different angular velocity is discussed. This concept is analytically and numerically demonstrated  in the case of cylinders, showing that generalized SCT operates efficiently in making rotating objects appear static to an external observer. Our proposal extends the realm of SCT, and brings it one step closer to its practical realization that involves moving objects. 
\end{abstract}

\date{\today}

\maketitle


\section{Introduction}


Inspired by the concept of photonic crystals \cite{yablonovitch1987inhibited,john1987strong,meade1995photonic,benisty2008photonic} and photonic crystal fibres \cite{knight1996all,russell2003photonic,white2002multipole,zollafoundations}, a new class of acoustic materials has emerged during the 1990s. These so-called phononic crystals (PCs), consist of a periodic arrangement of at least two materials with different densities \cite{kushwaha1997stop,vasseur1998experimental,tanaka1998surface,tanaka2000band}. These materials were shown to possess a frequency range over which sound wave propagation is prohibited (phononic bandgap) \cite{kushwaha1997stop,vasseur1998experimental}. These forbidden bands can also result in singular properties of sound waves, e.g., negative refraction at the interface between a classical medium and a phononic crystal \cite{zhang2004negative,yang2008membrane,liang2012tunable}, ultrasound tunneling \cite{yang2002ultrasound}, or tunable filtering and demultiplexing \cite{pennec2004tunable}, to cite a few \cite{cai2008acoustical,amin2015acoustically,cummer2016controlling,landi2018acoustic,wu2018perspective,assouar2018acoustic}. These PCs are difficult to miniaturize, as the bandgap appears at wavelengths in the same order as the period of the PCs, meaning a low frequency bandgap requires a large PC \cite{deymier2013acoustic,craster2012acoustic}. To overcome this major hurdle, active research has been directed towards other concepts involving local resonances \cite{liu2005analytic,di2018acoustic}, that best operate when the periodicity is much smaller than the wavelength (this is the so-called quasi-static, or long-wavelength limit). For instance, composite structures formed by locally resonating meta-atoms, so-called metamaterials (MMs) made their appearance at the turn of the century, both for electromagnetic \cite{pendry1996extremely,pendry1999magnetism,pendry2000negative,smith2000composite} and acoustic \cite{liu2000locally} waves. Based on analogies drawn from optical MMs, an acoustic MM consists of a heterostructure formed of resonant inclusions having characteristic dimensions smaller than the wavelength of the wave propagating in the medium, and vibrating on their natural modes of resonance \cite{li2004double}.

Controlling the propagation of waves using these engineered MMs is thus a considerable opportunity \cite{papanicolaou2012wave,farhat2016transformation}. For example, one can protect buildings from seismic waves \cite{farhat2014platonic,brule2020emergence} or tsunamis \cite{farhat2008broadband,dupont2015numerical,park2019hydrodynamic,zou2019broadband} by means of a large scale metamaterial which may guide the acoustic/elastic energy out of the area to be protected and may considerably attenuate the amplitude of the impinging waves. Defense applications are potentially very important as well, with the possibility, for example, of fabricating stealth systems (invisibility cloaks) \cite{farhat2016transformation}. The term "invisibility cloak" designates a coating whose material parameters, determined by the optical transformation process, make it possible to deflect any electromagnetic or elastodynamic wave \cite{pendry2006controlling,leonhardt2006optical,zolla2007electromagnetic}. If one places an object in the isolated interior area, then no incident wave can interact with this object since the cloak detours wave trajectories around it in such a way that for any external observer the field appears to be undisturbed. In other words, the object is both undetectable (invisible) and protected. Note that the concept of invisibility is distinct from that of stealth \cite{bahret1993beginnings}. The primary purpose of a stealth coating is to cancel the reflection coefficient in certain directions (typically those of a detection antenna). To do this, the idea is to absorb the incident waves or to reflect them to another direction. Interestingly, some species of moths have acquired dynamic acoustic camouflaging features thanks to some microstructure reminiscent of metamaterial surfaces \cite{neil2020thoracic}. Conversely, in an invisible device, one cancels both the reflection coefficient and the absorption, and one makes the transmission coefficient ideally unitary. The object included in the coating then has a zero electromagnetic/acoustic size and it has no "shadow" \cite{farhat2016transformation}. Other cloaking strategies were subsequently proposed through homogenization \cite{cai2007optical,farhat2008achieving,ergin2010three}, and/or scattering cancellation technique (SCT) \cite{alu2005achieving,chen2012invisibility}, and even suggested for other types of waves \cite{chen2007acoustic,dupont2011numerical,xu2015molding}.

All the above-mentioned devices and techniques operate for static objects, i.e., at rest. For instance, moving or rotating objects possess intrinsically different scattering signature \cite{graham1969effect,censor1971scattering,schoenberg1973elastic,censor1973two,peng2012lumped,lavery2013detection,farhadi2018acoustic,ramaccia2019phase,mazor2019rest,zhao2020acoustic} and require special treatment \cite{steinberg2006two,novitski2014finite}. Some intriguing applications were put forward with spinning elements such as gyroscopes \cite{steinberg2005rotating,steinberg2007splitting} and waveguide rotation sensors \cite{novitski2014finite}. In order to realize efficient cloaking devices for such rotating devices, one first needs to characterize the scattering response from these acoustic objects and then analyze the feasibility and physical difference of cloaking mechanism. A recent study for example considered cloaking structures that are moving in a rectilinear way, using spatiotemporal properties to counter the Doppler effect \cite{ramaccia2017doppler,ramaccia2018nonreciprocity}. A theory of a space-time cloak was also proposed to make a time interval undetectable for an observer \cite{mccall2010spacetime} and further demonstrated experimentally in a highly dispersive optical fibre \cite{fridman2012demonstration}. However, cloaking rotating objects was not achieved in any context. In the present paper, the use of the scattering cancellation technique \cite{guild2010cancellation,chen2012invisibility,farhat2012frequency,fleury2013quantum,farhat2016cloaking} to render spinning objects invisible for acoustic waves is proposed.

The rest of the paper is organized as follows. In the background and problem setup section, the equations of motion of a spinning acoustic object and its dispersion relation, that permit the study of acoustic scattering from multilayered spinning structures are put forward. In the following section, using Bessel expansions of the pressure fields, it is shown the possibility of cancelling the leading scattering orders from spinning objects by coating them with shells of tailored spinning velocity. In this section the effect of geometrical parameters and spinning velocity amplitude, on the scattering reduction is also analyzed. Finally, the obtained results are summarized in the concluding remarks.

\section{Background and Problem Setup}
\label{sec:section2}

\subsection{Acoustic equation in rotating media}

Let us consider time-harmonic waves, with dependence upon time $t$ proportional to $e^{-i\omega t}$, where $\omega$ is the angular wave frequency. One also assumes structures with cylindrical symmetry (invariance) [See Fig.~\ref{fig:fig_scheme}(a),] i.e., proportional to $e^{in\theta}$ so that derivatives in the azimutal direction $\theta$ produce terms of the form $in$, where $n$ is an integer and $i^2=-1$. 
One starts by invoking the mass and momentum conservation laws of acoustics and one expresses them in the laboratory frame of reference \cite{morse1968ku}. This is done by using the Eulerian specification of the flow field, i.e. by replacing time derivatives with material derivatives, i.e., $\partial_{t'}(\cdot)\rightarrow\partial_t(\cdot)+{\bf u}\cdot\nabla(\cdot)$, where the $'$ denotes derivatives taken in the rest frame of the moving fluid. Also, ${\bf u}={\bf u}_0+{\bf v}$ is the total velocity of the flow, whereas ${\bf u}_0$ is the bulk velocity, and ${\bf v}$ is the acoustical perturbation velocity. This results in modified conservation equations (See Appendix~\ref{sec:appendix}). 

One combines the modified equations, i.e., Eqs.~(\ref{eq:eqa3})-(\ref{eq:eqa4}) of the Appendix, by writing down each component [for instance, Eq.~(\ref{eq:eqa3}) contains two components, while Eq.~(\ref{eq:eqa4}) contains a single component.] Then, one uses ${\bf u}_0\cdot{\bf \nabla}=\Omega_1\partial/\partial\theta$ and one linearizes the equations, by keeping only first order quantities of acoustic perturbations, e.g., terms such as $({\bf v}\cdot\nabla){\bf v}$ are neglected, as shown in Appendix~\ref{sec:appendix} \cite{morse1968ku}. One thus obtains a linear system (of order 3) in terms of the variables $p_1$, $v_{r,1}$, and $v_{\theta,1}$ (assuming that the $z$-components of the fields are zero, as it is assumed in this first example, infinitely extended cylinders in the $z$-direction). This system of coupled partial differential equations (PDEs) can be expressed, in cylindrical coordinates, using the differential operator $\tilde{D}$, as $\tilde{D}\left(v_{r,1},v_{\theta,1},p_1\right)^T=0$, with $(\cdot)^T$ denoting the transpose of the vector in parentheses, i.e.,
\begin{equation}
   \begin{pmatrix}
  \zeta_{n,1} & -2\Omega_1 & \rho_1^{-1}\partial_r\\
  2\Omega_1 & \zeta_{n,1} & \left(\rho_1 r\right)^{-1}in\\
  r^{-1}+\partial_r & r^{-1}in & \rho_1^{-1}c_1^{-2}\zeta_{n,1}  
 \end{pmatrix}\begin{pmatrix}
  v_{r,1} \\
  v_{\theta,1}\\
  p_1  
 \end{pmatrix}={\bf 0}\; .
\label{eq:eq1}
\end{equation}
In this coupled differential system, one denotes $\partial_r=\partial/\partial r$ and the modified angular frequency $\zeta_{n,1}=i(n\Omega_1-\omega)$. From the system of Eq.~(\ref{eq:eq1}), one derives the equation verified by the pressure $p_1$, in the cylindrical coordinates, that is
\begin{equation}
\frac{\partial^2 p_1}{\partial r^2}+\frac{1}{r}p_1+\left(\beta_{n,1}^2-\frac{n^2}{r^2}\right)p_1=0\, ,
\label{eq:eq2}    
\end{equation}
which is the Helmholtz equation expressed in cylindrical coordinates, assuming an effective wavenumber
\begin{equation}
\beta_{n,1}=\sqrt{\frac{-\left(4\Omega_1^2+\zeta_{n,1}^2\right)}{c_1^2}}\, .
\label{eq:eq3}    
\end{equation}
One verifies that for $\Omega_1=0$, one recovers the classical dispersion $\beta_1=\omega/c_1$, with $c_1=\sqrt{\kappa_1/\rho_1}$ the speed of sound inside the object (all parameters related to the object are denoted with subscript 1 and those related to free-space with subscript 0). The behavior of $\beta_{n,1}$, i.e. the spinning effective wavenumber is depicted in Fig.~\ref{fig:fig_scheme}(b). As the parameter $\zeta_{n,1}$ is complex, $\beta_{n,1}$ possesses both imaginary and real parts (i.e. damping). For example, for $\beta_{0,1}$ (i.e. $n=0$,) in the domain $|\alpha_1|=|\Omega_1/\omega|\leq1/2$, only the real part exists and decays exponentially, while reaching 0 for $\alpha_1=\pm1/2$. The imaginary part is zero in this domain and increases quasi-linearly. The orders $n=\pm1$ possess similar and symmetric behavior. The orders $n=\pm2$ have slightly different behavior, which is not symmetric with respect to $\alpha_1$. It should be mentioned that for $\alpha_1=0$, i.e. no spinning, both wavenumbers ($\beta_{n,1}$ and $\beta_1$) are equal, as expected. In the domain $|\alpha_1|\ll1$, one should expect no damping of the spinning wavenumbers, as observed. 

Now, Eq.~(\ref{eq:eq2}) shall be complemented with adequate boundary conditions. In the case of media at rest, one has continuity of the pressure field $p_1$ and the normal component of the velocity field (proportional to the displacement field) $v_{r,1}\propto p_1/\rho_1$. In the case of spinning media, one has continuity of the pressure and of the normal displacement $\psi_{r,1}$ [See Eq.~(\ref{eq:eqa5}) in Appendix~\ref{sec:appendix}] \cite{morse1968ku},
\begin{eqnarray}
\nonumber
\psi_{r,1}&=&\frac{\zeta_{n,1} v_{r,1}+\Omega_1 v_{\theta,1}}{\zeta_{n,1}^2+\Omega_1^2}\\
&=&\frac{\left(2\Omega_1^2-\zeta_{n,1}^2\right)\partial_r p_1-3i\zeta_{n,1}\Omega_1 np_1/r}{\rho_1\left(4\Omega_1^2+\zeta_{n,1}^2\right)\left(\Omega_1^2+\zeta_{n,1}^2\right)}\, .
\label{eq:eq4}    
\end{eqnarray}
By letting $\Omega_1=0$ in Eq.~(\ref{eq:eq4}), on gets a displacement proportional to $(1/\rho_1)\partial_rp_1$ as in the case for acoustic waves in media at rest.

\subsection{Bessel expansion and scattering from bare spinning objects}
Let us now turn to the main problem of characterizing the scattering from rotating cylindrical objects, at uniform angular velocity $\Omega_1$. First one considers a bare cylindrical object of radius $r_1$ rotating in free-space with density and bulk modulus $\rho_1$ and $\kappa_1$, respectively. At this stage, one will derive the general equation for any properties of the rotating object, and later, it will be assumed that $\rho_1=\rho_0$ and $\kappa_1=\kappa_0$ to single out the pure effects due to spinning. An acoustic plane-wave of amplitude 1 is incident on the structure. For simplicity and without loss of generality, let us assume that the wave is in the $x-y$ plane, and that it propagates in the $x$-direction. It can thus be expressed as $p^\textrm{inc}=e^{i{\bf \beta_0}x}=e^{i\beta_0 r\cos\theta}$, by ignoring the time-harmonic dependence, for now. The expansion of this incident plane wave in terms of Bessel functions takes the form
\begin{equation}
p^\textrm{inc}=\sum_{-\infty}^{+\infty}i^nJ_n\left(\beta_0r\right)e^{in\theta}\, .
\label{eq:eq5}    
\end{equation}
The scattered field is expanded in terms of Hankel functions of the first kind, to ensure that the Sommerfield radiation condition is satisfied, i.e.
\begin{equation}
p^\textrm{scat}=\sum_{-\infty}^{+\infty}i^ns_nH_n^{(1)}\left(\beta_0r\right)e^{in\theta}\, ,
\label{eq:eq6}    
\end{equation}
for $r>r_1$ and with $s_n$ the scattering coefficients to be determined using the boundary conditions at the interfaces of the structure. Hence, the field in region 0 is $p_0=p^\textrm{inc}+p^\textrm{scat}$. These scattering coefficients intervene in the definition of the scattering amplitude $f(\theta)\propto\sqrt{r}\lim_{x\to\infty} p^\textrm{scat}(r,\theta)$, which is a measure of the acoustic scattering strength in the direction $\theta$. The total scattering cross-section (SCS) is the integration over all angles $\theta$ of the scattering amplitude and represents a scalar measure of the total scattering (irrespective of direction), and in the two-dimensional (2D) scenario is proportional to a length. For instance, one has
\begin{equation}
\sigma^\textrm{scat}=\frac{4}{\beta_0}\sum_{-\infty}^{+\infty}|s_n|^2\, .
\label{eq:eq7}    
\end{equation}
To complete the expansion of the pressure fields, one considers now the case of the spinning disc of radius $r_1$ that is different from scattering objects that were considered in previous studies, so far. In this case and owing to the previous results, the pressure field in the region $r\leq r_1$ is given by
\begin{equation}
p_1=\sum_{-\infty}^{+\infty}i^na_nJ_n\left(\beta_{n,1}r\right)e^{in\theta}\, ,
\label{eq:eq8}    
\end{equation}
with $\beta_{n,1}$ given in Eq.~(\ref{eq:eq3}) and $a_n$ unknown coefficients to be determined by the boundary conditions along with $s_n$. Now by equating the pressure and the displacement [See Eq.~(\ref{eq:eq4})] at the boundary $r=r_1$, i.e.
\begin{widetext}
\begin{eqnarray}
\nonumber
p^\textrm{inc}\left(r_1\right)+p^\textrm{scat}\left(r_1\right)&=&p_1\left(r_1\right)\, ,\\
\frac{1}{\rho_0\omega^2}\left.\frac{\partial\left(p^\textrm{inc}+p^\textrm{scat}\right)}{\partial r}\right |_{r=r_1}&=&\left.\frac{\left(2\Omega_1^2-\zeta_{n,1}^2\right)\partial_r p_1-3i\zeta_{n,1}\Omega_1 np_1/r}{\rho_1\left(4\Omega_1^2+\zeta_{n,1}^2\right)\left(\Omega_1^2+\zeta_{n,1}^2\right)}\right |_{r=r_1}\, .
\label{eq:eq9}    
\end{eqnarray}
Equation~(\ref{eq:eq9}) yields with the previous expansions a set of linear systems, for each azimutal order $n$, thanks to the orthogonality of the functions $e^{in\theta}$, i.e.
\begin{equation}
   \begin{pmatrix}
  J_n\left(\beta_{n,1}r_1\right) & -H_n^{(1)}\left(\beta_0r_1\right)\\
  \Pi_{J_n} & -\frac{\beta_0}{\omega^2\rho_0}H_n^{(1)'}\left(\beta_0r_1\right)
  \end{pmatrix}\begin{pmatrix}
  a_n \\
  s_n  
 \end{pmatrix}=\begin{pmatrix}
  J_n\left(\beta_0r_1\right) \\
  \frac{\beta_0}{\omega^2\rho_0}J_n'\left(\beta_0r_1\right)  
 \end{pmatrix}\; ,
\label{eq:eq10}
\end{equation}
where the coefficient $\Pi_{J_n}$ is expressed as
\begin{equation}
\Pi_{J_n}=\frac{\left(2\Omega_1^2-\zeta_{n,1}^2\right)\beta_{n,1}J_n'\left(\beta_{n,1}r_1\right)-\frac{3\zeta_{n,1}\Omega_1 in}{r_1} J_n\left(\beta_{n,1}r_1\right)}{\rho_1\left(4\Omega_1^2+\zeta_{n,1}^2\right)\left(\Omega_1^2+\zeta_{n,1}^2\right)}\, .
\label{eq:eq11}    
\end{equation}

Equation~(\ref{eq:eq11}) shows clearly for the specific case of scattering from spinning objects, that the multipoles of orders $n$ and $-n$ give different contributions. The scattering coefficient $s_n$ can be easily obtained from Eq.~(\ref{eq:eq10}), i.e.
\begin{equation}
s_n=
\begin{vmatrix}
  J_n\left(\beta_{n,1}r_1\right) & J_n\left(\beta_0r_1\right)\\
  \Pi_{J_n} & \frac{\beta_0}{\omega^2\rho_0}J_n'\left(\beta_0r_1\right)
  \end{vmatrix}
   {\begin{vmatrix}
  J_n\left(\beta_{n,1}r_1\right) & -H_n^{(1)}\left(\beta_0r_1\right)\\
  \Pi_{J_n} & -\frac{\beta_0}{\omega^2\rho_0}H_n^{(1)'}\left(\beta_0r_1\right)
  \end{vmatrix}}^{-1}\; ,
\label{eq:eq12}
\end{equation}
\end{widetext}
where $|M|$ denotes the determinant of a matrix $M$.

In order to single out the effect of rotation on the scattering, one considers an object with the same density and bulk modulus as the surrounding environment, i.e. $\rho_1=\rho_0$ and $\kappa_1=\kappa_0$. This leaves us with only the rotation angular velocity $\Omega_1$ of the object ($r\leq r_1$). A scenario of interest is that of small objects compared to the sound wavelength, i.e. $\beta_{0}r_1\ll 1$ and $\beta_{n,1}r_1\ll 1$. The first multipole terms are thus given by
\begin{eqnarray}
\nonumber
s_0&=&\frac{3i\pi}{4}\frac{\alpha_1^2\left(\beta_1r_1\right)^2}{1-\alpha_1^2}+O\left(\left(\beta_1r_1\right)^4\right)\, ,\\
\nonumber
s_{\pm1}&=&\frac{i\pi}{4}\frac{\alpha_1\left(\beta_1r_1\right)^2}{\pm2+\alpha_1}+O\left(\left(\beta_1r_1\right)^4\right)\, ,\\
\nonumber
s_{\pm2}&=&\frac{i\pi}{32}\frac{\alpha_1\left(\beta_1r_1\right)^4}{\mp2+\alpha_1}+O\left(\left(\beta_1r_1\right)^6\right)\, ,\\
s_{\pm n}&=&f_{\pm n}\left(\alpha_1\right)\left(\beta_1r_1\right)^{2n}+O\left(\left(\beta_1r_1\right)^{2n+2}\right)\, .
\label{eq:eq13}    
\end{eqnarray}
In Eq.~(\ref{eq:eq13}), $f_\pm$ denote functions of the variable $\alpha_1$. The upper/lower sign in the second and third lines correspond to the positive/negative coefficient, respectively. Also, $O(\cdot)$ denotes the Landau notation (of a function of the same order) \cite{landau2000handbuch}. It may be noted that, if the angular rotation velocity of the fluid goes to zero, all the scattering orders $s_n$ vanish without exception. A case of interest is that of small rotation angular velocity, so the denominators in Eq.~(\ref{eq:eq13}) are close to 1 and can be omitted, thus one has $s_0\propto\alpha_1^2\omega^2$, $s_{\pm1}\propto\alpha_1\omega^2$, and $s_{\pm2}\propto\alpha_1\omega^4$. In classical scattering from non-rotating acoustic objects (or 2D electromagnetism), it is well known that the scattering cross-section is dominated by both the zeroth-order and first-order, i.e. the monopole $s_0$ and the dipole $s_1$ \cite{wu2006effective}. However, from Eq.~(\ref{eq:eq13}), one can see that $s_0/s_{\pm1}\propto\alpha_1\ll 1$ and $s_{\pm2}/s_{\pm1}\propto\omega^2\ll 1$. Hence, unlike for the case of acoustics at rest \cite{farhat2012frequency}, the SCS of spinning objects is dominated by the dipole terms $s_{\pm{1}}$. The higher order terms scale as $(\beta_1r_1)^{2n}$ and do not contribute significantly to the scattering, although in Eq.~(\ref{eq:eq11}) one has terms proportional to $n$. However, the peculiar behavior of Bessel functions makes the higher order multipoles negligible in the quasi-static limit.

Another interesting remark about scattering of spinning fluids can be immediately seen upon inspection of Eq.~(\ref{eq:eq13}). One can see that the scattering coefficients possess poles for determined values of $\alpha_1$. Namely, these are $\omega=\pm\Omega_1$ for $s_0$, $\mp\Omega_1/2$ for $s_{1,-1}$ and $\pm\Omega_1/2$ for $s_{2,-2}$. Thus for these frequencies, resonant scattering may be observed. For instance,
\begin{equation}
\sigma^\textrm{scat}\approx\frac{\pi}{2}\left(\beta_0r_1\right)\frac{\alpha_1^2\left(\alpha_1^2+4\right)}{\left(\alpha_1^2-4\right)^2}\, .
\label{eq:eq14}
\end{equation}

\begin{figure}
    \centering
    \includegraphics[width=0.85\columnwidth]{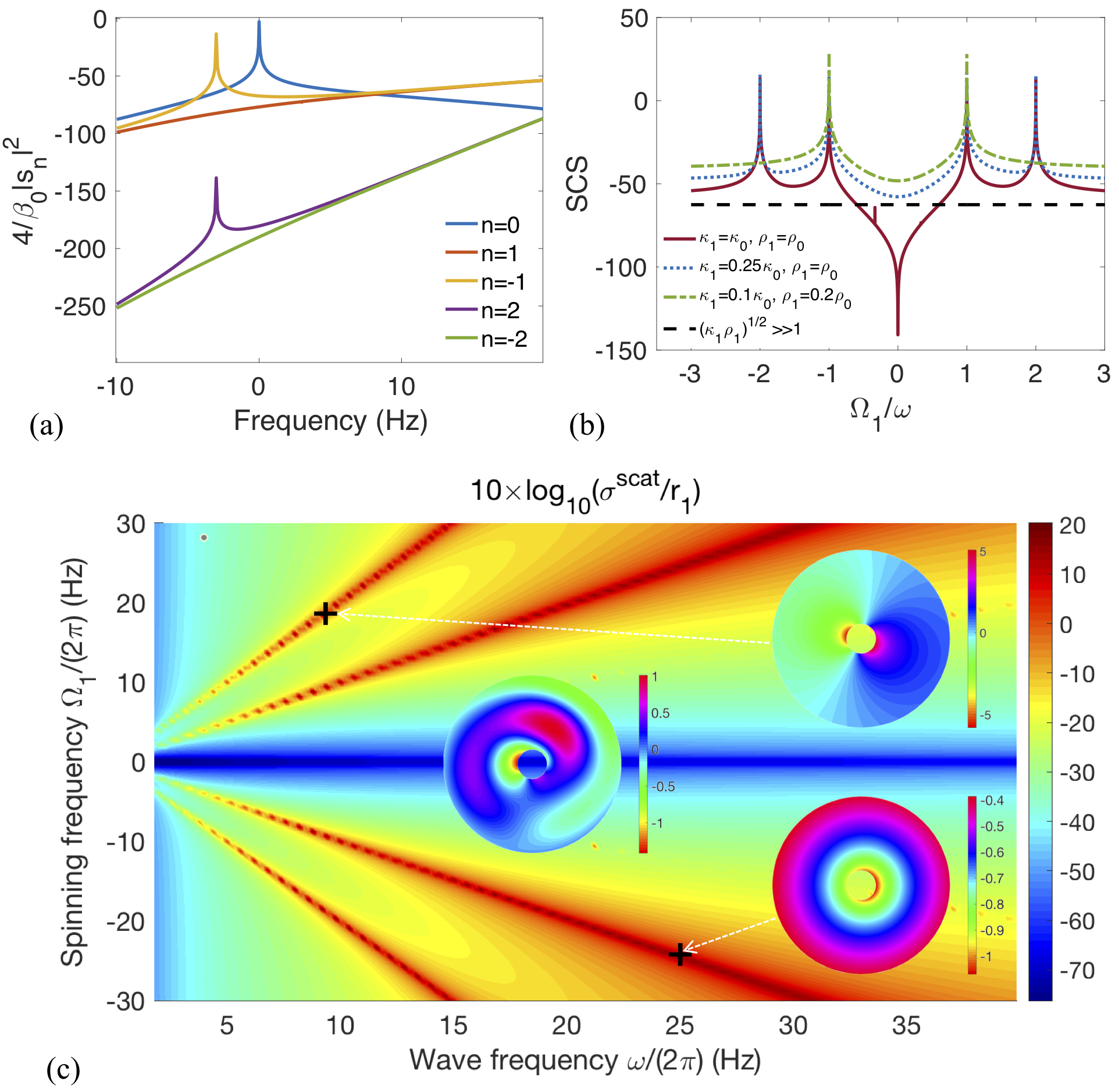}
    \caption{(a) Scattering coefficients $4/\beta_0|s_n|^2$ in logarithmic scale versus frequency (in logarithmic scale, too) for a spinning object ($\Omega_1/(2\pi)=1$ Hz) of radius $r_1=1$ m, bulk modulus and density $\kappa_1=\kappa_0=2.22$ GPa and $\rho_1=\rho_0=10^3\textrm{kg}/\textrm{m}^3$, respectively, for $n=0,\pm1,\pm2$. (b) Total normalized SCS ($\sigma^\textrm{scat}/r_1$) with 21 scattering orders taken into account ($n=-10:10$) versus the normalized spinning velocity for different kinds of objects, ranging from soft ($\sqrt{\kappa_1\rho1}\ll1$), "normal" ($\sqrt{\kappa_1\rho1}\approx1$), to hard-wall (See Section~\ref{sec:appendixb}) ($\sqrt{\kappa_1\rho1}\gg1$). (c) Normalized SCS in logarithmic scale, of the spinning cylinder vs the frequency of the acoustic wave $\omega$ and the spinning velocity $\Omega_1$ for the same physical parameters as the environment (water). The inset plots $\Re{(p_0)}$ at corresponding parameters. The middle one corresponds to high frequency $\Omega_1=2\omega=2\pi\times500$ Hz.}
    \label{fig:fig_sn_bare}
\end{figure}

Figure.~\ref{fig:fig_sn_bare}(a) plots these normalized scattering coefficients $4/\beta_0|s_n|^2$ in logarithmic scale versus frequency (in logarithmic scale, too) for a spinning object (made of water, as the surrounding, and separated from it by a thin membrane), with $\Omega_1=2\pi$ rad, of radius $r_1=1$ m, bulk modulus and density $\kappa_1=\kappa_0=2.22$ GPa and $\rho_1=\rho_0=10^3\textrm{kg}/\textrm{m}^3$, respectively, for $n=0,\pm1,\pm2$. These plots show that although the object has the same physical parameters as the environment (water, here, for instance) resonant modes take place at specific frequencies given by Eq.~(\ref{eq:eq13}). It should be also noted that both modes $n=0$ and $n=\pm1$ dominate, as can be anticipated from Eq.~(\ref{eq:eq13}). Also the resonance of modes $n=1$ and $n=-2$ cannot be seen here as one uses positive $\Omega_1(=2\pi$ rad). Figure.~\ref{fig:fig_sn_bare}(b) depicts the total scattering cross-section $\sigma_\textrm{scat}$ with 21 scattering orders taken into account ($n=-10:10$) versus the normalized spinning velocity for different kinds of objects, ranging from soft, i.e. $\sqrt{(\kappa_1\rho_1)/(\kappa_0\rho_0)}\ll1$ [green line in Fig.~\ref{fig:fig_sn_bare}(b)] "non-rigid", i.e. $\sqrt{(\kappa_1\rho_1)/(\kappa_0\rho_0)}\approx1$ [red and blue lines in Fig.~\ref{fig:fig_sn_bare}(b)], to hard-wall (rigid) (detailed in Section~\ref{sec:appendixb}), i.e. $\sqrt{(\kappa_1\rho_1)/(\kappa_0\rho_0)}\gg1$ [black dashed line in Fig.~\ref{fig:fig_sn_bare}(b)]. The resonant scattering can be seen from all these objects around the predicted spinning velocities. Here the frequency is fixed at 1 Hz (quasistatic limit). It should be noted that the presence of these Mie resonances is unusual in acoustics, where homogeneous objects do not possess low frequency resonance. The only case of low-frequency Mie resonances, concerns flexural waves scattered off thin-plate objects as was analyzed in Ref. \cite{farhat2017localized}, originating from the peculiar nature of flexural biharmonic waves obeying a fourth order PDE \cite{graff2012wave}. However, in the present scenario, these resonances re due to pure rotation. Figure~\ref{fig:fig_sn_bare}(c) plots contours of normalized SCS $\sigma^\textrm{scat}/r_1$ (in logarithmic scale) of a scatterer with the same physical properties as the surrounding (water) for varying frequencies $\omega$ and spinning speeds $\Omega_1$. This plot clearly shows that the SCS has two resonances (marked with dark red color) for each spinning speed. Moreover, the blue horizontal thick linear region at the center with blue color (i.e. zero scattering) corresponds to very low down to zero spinning speeds, and as the scatterer possesses the same density and bulk modulus of the surrounding, it does not scatter at all at these low spinning speeds. On the other hand, if one takes vertical cuts along this 2D graph, four resonances occur (in a symmetric manner with respect to $\Omega_1$) as it transpires from Fig.~\ref{fig:fig_sn_bare}(b) and as predicted from Eq.~(\ref{eq:eq13}.)

The inset of Fig.~\ref{fig:fig_sn_bare}(c) plots the real part distribution of the pressure field [$\Re(p)$] in the scattering region (i.e. region 0) in the presence of the spinning acoustic cylinder. These plots correspond to frequencies and spinning speeds of different resonating modes, as depicted in Fig.~\ref{fig:fig_sn_bare}(c). It may be seen that the pressure field takes very large values (in comparison to the non-resonating case, where $\Re(p)\approx 10^{-3}p_0$) which is coherent with the observed Mie resonance due to the spinning fluid.

\section{Scattering Cancellation for Spinning Cylinders}
\label{sec:section3}

\subsection{Bessel function expansion for the core-shell structure}

Let us now turn to the analysis of cloaking the spinning objects using the paradigm of the scattering cancellation technique (SCT) \cite{alu2005achieving}. One considers a core-shell structure, depicted in Fig.~\ref{fig:fig_scheme}(a), with an object of radius $r_1$ and a shell of radius $r_2$. The parameters of the object and the shell are denoted by $\rho_{1,2}$, $\kappa_{1,2}$, and $\Omega_{1,2}$, for the density, bulk modulus, and angular velocity, respectively. On the other side, the parameters of free-space are just $\rho_0$ and $\kappa_0$, as the fluid in region 0 is at rest. The field expansions are similar to the ones of a bare object. However, one has now an additional domain (the shell) $r_1<r\leq r_2$, where the pressure field can be expanded as
\begin{equation}
p_2=\sum_{-\infty}^{+\infty}i^n\left[b_nJ_n\left(\beta_{n,2}r\right)+c_nY_n\left(\beta_{n,2}r\right)\right]e^{in\theta}\, ,
\label{eq:eq15}    
\end{equation}
with $Y_n$ the Bessel function of second kind, $\beta_{n,2}=\sqrt{-(4\Omega_2^2+\zeta_{n,2}^2)/c_2^2}$ and $c_2=\sqrt{\kappa_2/\rho_2}$.

The obtained scattering system for this structure is thus obtained by applying the same boundary conditions at the interfaces $r=r_1$ and $r=r_2$, taking into account that the fluid is either rotating or at rest. This leads to
\begin{widetext}
\begin{equation}
   \begin{pmatrix}
  0 & J_n\left(\beta_{n,2}r_2\right) & Y_n\left(\beta_{n,2}r_2\right) & -H_n^{(1)}\left(\beta_0r_2\right)\\
  0 & \Pi_{J_n}\left(\beta_{n,2}r_2\right) & \Pi_{Y_n}\left(\beta_{n,2}r_2\right) & -\frac{\beta_0}{\omega^2\rho_0}H_n^{(1)'}\left(\beta_0r_2\right)\\
  -J_n\left(\beta_{n,1}r_1\right) & J_n\left(\beta_{n,2}r_1\right) & Y_n\left(\beta_{n,2}r_1\right) & 0\\
  -\Pi_{J_n}\left(\beta_{n,1}r_1\right) & \Pi_{J_n}\left(\beta_{n,2}r_1\right) & \Pi_{Y_n}\left(\beta_{n,2}r_1\right) & 0
  \end{pmatrix}\begin{pmatrix}
  a_n\\
  b_n\\
  c_n\\
  s_n 
 \end{pmatrix}=\begin{pmatrix}
  J_n\left(\beta_0r_2\right)\\
  \frac{\beta_0}{\omega^2\rho_0}J_n'\left(\beta_0r_2\right)\\
  0\\
  0
 \end{pmatrix}\; ,
\label{eq:eq16}
\end{equation}
\end{widetext}
with the functionals $\Pi_{Y_n}$ given in the same way as $\Pi_{J_n}$, shown in Eq.~(\ref{eq:eq11}), up to the replacement of $J_n$ by $Y_n$. The scattering coefficient is thus $s_n=|M|/|\tilde{M}|$, where $M$ is the $4\times4$ matrix in the LHS of Eq.~(\ref{eq:eq16}) and $\tilde{M}$ is the matrix obtained from $M$ by replacing its fourth column vector by the vector in the RHS of Eq.~(\ref{eq:eq16}). Solving Eq.~(\ref{eq:eq16}) is straightforward using a numerical software such as Matlab \cite{matlab}, and this will be performed later to characterize and analyze this peculiar cloaking mechanism. 

\subsection{Analysis of the SCT}

In order to gain more insight, and due to the general complexity of this linear system, it is instructive to analyze the long wavelength limit (as done for the bare object in previous section) corresponding to acoustically small objects and shells, i.e. $\beta_0r_{1,2}\ll1$ and $\beta_{n,1,n,2}r_{1,2}\ll1$. Note that with the values of the parameters in this study, it is sufficient to impose the first condition $\beta_0r_1\ll 1$. Under these assumptions, and by denoting $\Omega_{1}=\alpha_{1}\omega$, $\Omega_{2}=\alpha_{2}\omega$, $r_2=r_1/\gamma$, and by choosing without loss of generality $\rho_2=\rho_1=\rho_0$ and $c_2=c_1=c_0$, in order to single out the effect of spinning (by ignoring scattering due to the acoustic impedance mismatch due to inhomogeneities), one obtains for the leading scattering orders, as discussed in the previous sub-section, 
\begin{eqnarray}
\nonumber
s_0&=&\frac{3i\pi}{4\gamma^2}\frac{\left[\gamma^2\alpha_1^2-\left(-1+\gamma^2+\alpha_1^2\right)\alpha_1\alpha_2\right]}{\left(-1+\alpha_1^2\right)\left(-1+\alpha_1\alpha_2\right)}\left(\beta_0r_1\right)^2\\
&+&O\left(\left(\beta_0r_1\right)^4\right)\, ,
\label{eq:eq17}    
\end{eqnarray}

and 
\begin{equation}
s_{\pm1}=\frac{i\pi}{4\gamma^2}\frac{A_{\pm1}}{B_{\pm1}}\left(\beta_0r_1\right)^2+O\left(\left(\beta_0r_1\right)^3\right)\, ,    
\label{eq:eq18}
\end{equation}
with
\begin{eqnarray}
\nonumber
A_{\pm1} &=& \pm2\gamma^2\alpha_1+\alpha_2\left(\pm2\mp2\gamma^2+\alpha_1-6\gamma^2\alpha_1\right)\\
\nonumber
&+&\alpha_2^2\left(-3+6\gamma^2\pm\alpha_1\pm4\gamma^2\alpha_1\right)\\
 &+&\alpha_2^3\left(\pm1\mp4\gamma^2+6\gamma^2\alpha_1\right)+\alpha_2^4\left(6\left(1-\gamma^2\right)\right)\, ,
 \label{eq:eq19}
\end{eqnarray}
and
\begin{eqnarray}
\nonumber
B_{\pm1} &=& 4\pm2\alpha_1+\alpha_2\left(\mp4+3\alpha_1-2\gamma^2\alpha_1\right)\\
\nonumber
&+&\alpha_2^2\left(-1+2\gamma^2\pm\alpha_1\mp2\gamma^2\alpha_1\right)\\
 &+&\alpha_2^3\left(\pm13\pm2\gamma^2+6\gamma^2\alpha_1\right)+\alpha_2^4\left(6\left(1-\gamma^2\right)\right)\, .
 \label{eq:eq20}
\end{eqnarray}
In Eqs.~(\ref{eq:eq19})-(\ref{eq:eq20}) the upper (lower) sign correspond to the order $n=1$ ($n=-1$). Note that as $\alpha_2\rightarrow0$ i.e. the shell is at rest, the expressions of $s_0$ and $s_{\pm1}$ given in Eqs.~(\ref{eq:eq17})-(\ref{eq:eq20}) reduce to the ones given in Eq.~(\ref{eq:eq13}), as expected.

In order to cancel the total SCS, i.e. $\sigma^\textrm{scat}$, one has to enforce $s_0=0$ and $s_{\pm1}=0$. For small angular rotation speeds, only $s_{\pm1}$ is significant (as one has seen earlier from Eq.~(\ref{eq:eq13})). and it is safe to ignore the contribution of the higher order multipoles ($|n|\geq2$), as these scale with $\left(\beta_0r_1\right)^{2n}$ (their squared amplitude, i.e. their contribution to the SCS, from Eq.~(\ref{eq:eq7}) scales with $\left(\beta_0r_1\right)^{4n-1}$, which is even smaller). 

\begin{figure}[t!]
    \centering
    \includegraphics[width=0.55\columnwidth]{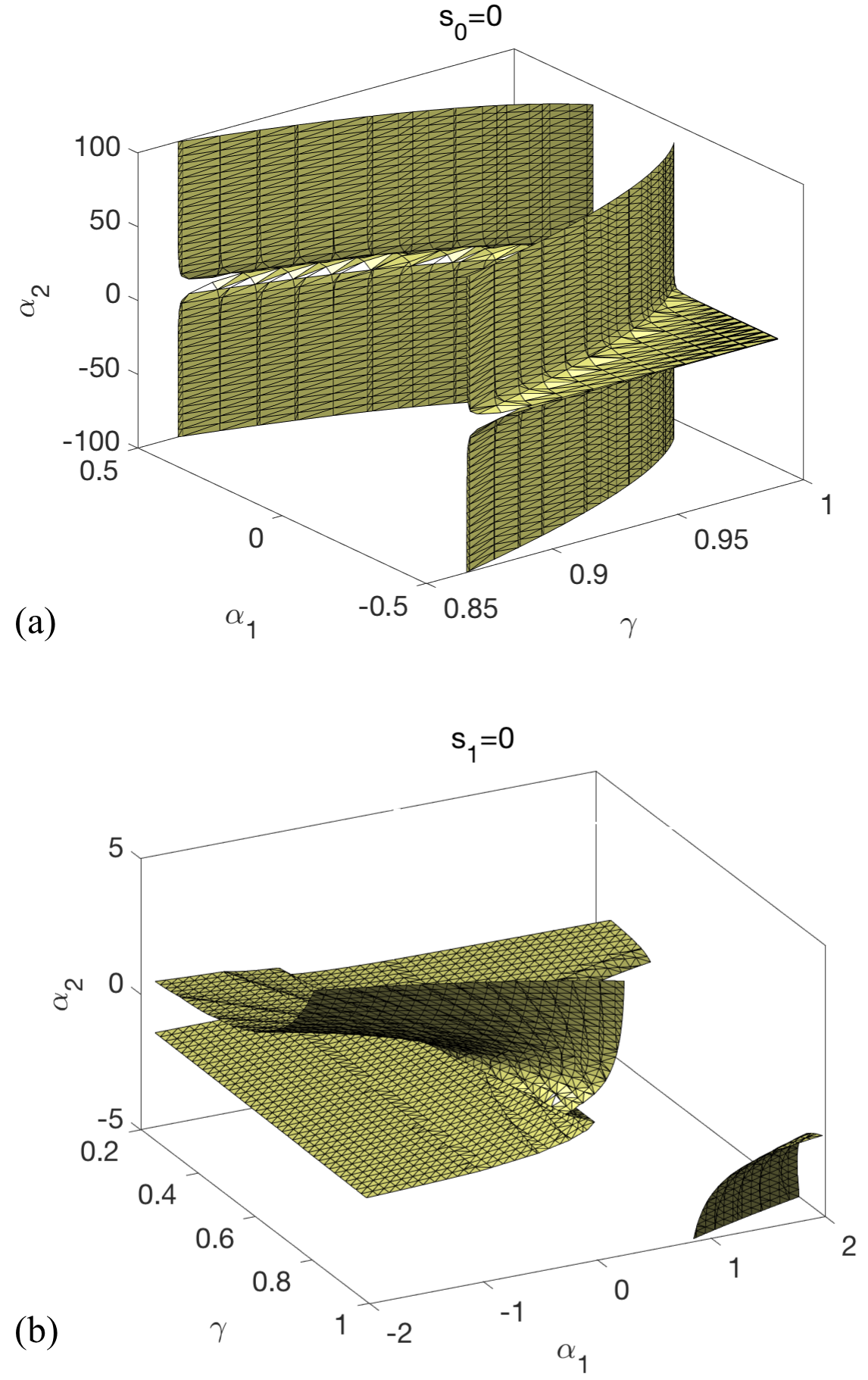}
    \caption{(a) Contour plot of $\alpha_2$ versus $\alpha_1$ and $\gamma$ for the first order ($n=0$) condition, given in Eq.~(\ref{eq:eq17}). (b) Contour plot of $\alpha_2$ versus $\alpha_1$ and $\gamma$ for the second order condition ($n=\pm1$), given in Eqs.~(\ref{eq:eq18})-(\ref{eq:eq20}).}
    \label{fig:fig_exact_cloaking}
\end{figure}

First, enforcing $|s_0|=0$, one derives the quasistatic condition of SCT, i.e.
\begin{equation}
\gamma^2\alpha_1^2-(-1+\gamma^2+\alpha_1^2)\alpha_2^2=0\, ,
\label{eq:eq21}    
\end{equation}
which relates $\alpha_2$, $\alpha_1$, $r_1$, and $r_2$ (via $\gamma$). It is found that, to satisfy Eq.~(\ref{eq:eq20}), $\alpha_2$ must take positive and/or negative values. Note that positive (resp. negative) angular velocity just means an anticlockwise (resp. clockwise) rotation. One thus has
\begin{equation}
\alpha_2=\pm\frac{\gamma\alpha_1}{\sqrt{(-1+\gamma^2+\alpha_1^2)}}\, .
\label{eq:eq22}    
\end{equation}
Also, when $\gamma^2+\alpha_1^2\leq1$, no solution can be abtained, that may cancel the scattering monopole $s_0$. The behavior of $\alpha_2$, versus $\gamma$ and $\alpha_1$, corresponding to Eq.~(\ref{eq:eq17}) is depicted in Fig.~\ref{fig:fig_exact_cloaking}(a). Specifically, one can observe that for the domain $\gamma^2+\alpha_1^2\leq1$, no solution for $\alpha_2$ can be obtained (empty region of the plot). For $\gamma^2+\alpha_1^2=1$, very high positive (and negative) values of $\alpha_2$ are required. On the other hand, when the condition on $\gamma^2+\alpha_1^2$ is relaxed, small values of $\alpha_2$ are sufficient. It should be also noted that $\alpha_2$ is symmetric with respect to the variation of $\alpha_1$ ($\alpha_1$ and $-\alpha_1$ give the same values of $\alpha_2$) as seen from Fig.~\ref{fig:fig_exact_cloaking}(a).

Let us now turn to the analysis of cancelling the leading scattering dipole orders $s_{\pm 1}$. In fact, Eq.~(\ref{eq:eq19}) is of fourth order, so one may expect to obtain four distinct solutions for $\alpha_2$. This is exactly what may be observed in Fig.~\ref{fig:fig_exact_cloaking}(b), where four branches can be distinguished in this three-dimensional contourplot. In this scenario, one can see that there is a lack of symmetry with respect to $\alpha_1$, due to the presence of the dipole order term in the equation ($n=\pm1$). Clockwise an anti-clockwise rotations $\Omega_2$ are thus viable ways to counteract the anti-clockwise rotation of the object and make it look static to external observers (by cancelling the $n=\pm1$ multipoles). The angular rotation speeds needed here are also comparable to the speed of the object to cancel. The graphs of Fig.~\ref{fig:fig_exact_cloaking} are only dependent of frequency through the parameters $\alpha_i=\Omega_i/\omega$.

\begin{figure}
    \centering
    \includegraphics[width=0.9\columnwidth]{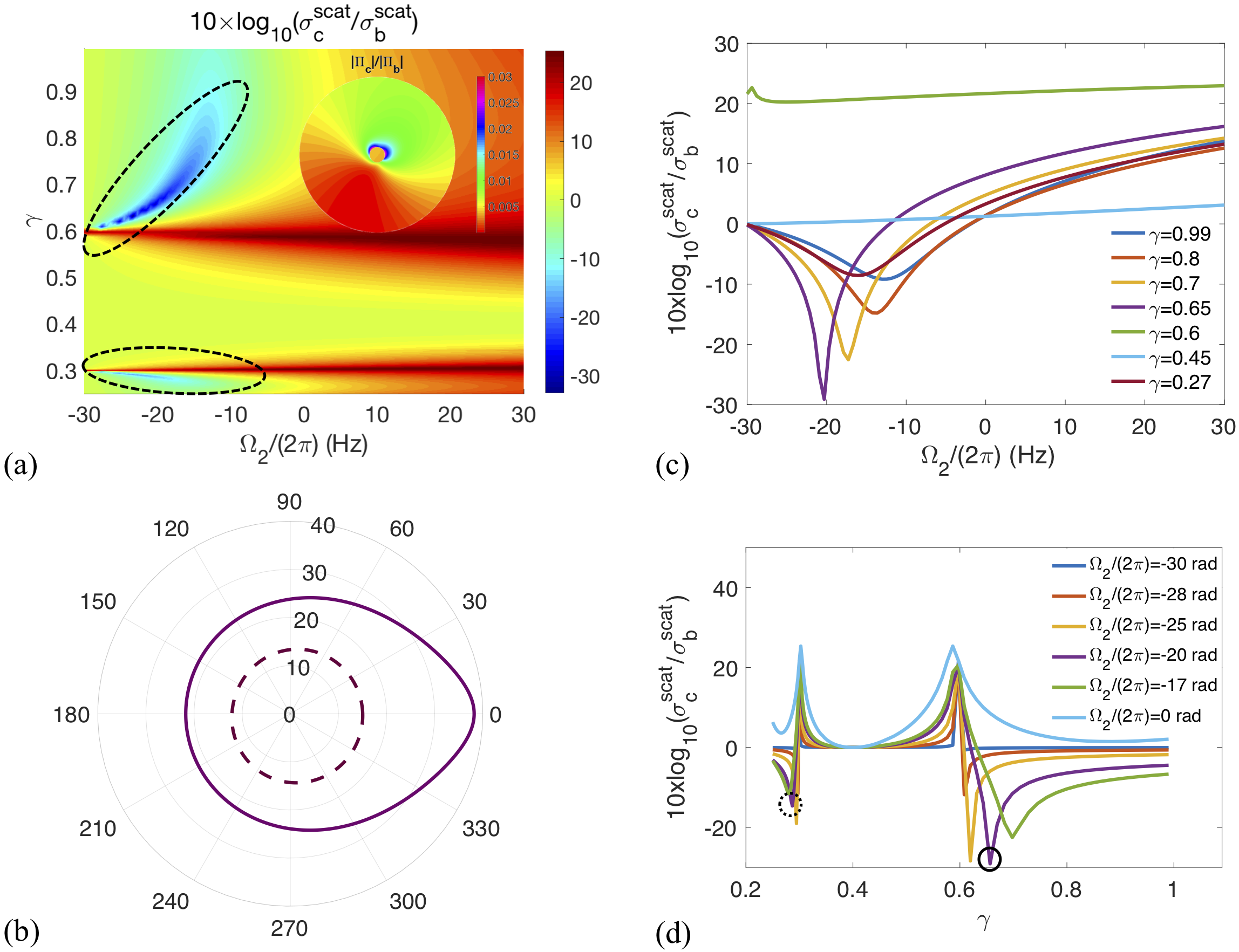}
    \caption{(a) Normalized SCS ($\sigma^\textrm{scat}_c/\sigma^\textrm{scat}_b$) in logarithmic scale, (where the subscripts c and b refer to the scattering cross-section of the obstacle and cloaked structure, respectively) versus the spinning frequency of the cloaking shell $\Omega_2/(2\pi)$ and the ratio $\gamma$. The highlighted region represents the locations of optimized scattering reduction, with a value exceeding 30 dB. The inset gives the acoustic Poynting vector of the cloaked structure normalized by the one of the bare object in the scattering region. (b) Scattering amplitude ($|f(\theta)|^2$) in logarithmic scale for the cloaked structure (normalized by the amplitude of the bare object) for $\Omega_2/(2\pi)=-20$ Hz two different radii  of the cloak, corresponding to the highlighted values from (d). (c) Normalized SCS versus the spinning frequency of the cloaking shell $\Omega_2/(2\pi)$ for various values of $\gamma$. (d) Normalized SCS versus $\gamma$ for various values of $\Omega_2/(2\pi)$. All these figures were plotted for a frequency of $\omega/(2\pi)=16.75$ Hz and $\Omega_1/(2\pi)=15$ Hz.}
    \label{fig:fig_scs_2d_cloak}
\end{figure}

Next, one considers the general case where one does not make use of asymptotic (quasistatic) approximation, and solve the exact scattering problem, stemming from Eq.~(\ref{eq:eq16}). The angular rotation speed of the fluid in region 1 ($r\leq r_1$) is $\Omega_1/(2\pi)=15$ Hz and its density and bulk modulus are assumed, as before, equal to those of free-space (water). Here, the frequency of the wave is chosen as $\omega/(2\pi)=16.75$ Hz (high spinning regime, i.e. $\Omega_1\approx\omega$). $\sigma^\textrm{scat}_c$ of the total object-shell structure is normalized with the SCS of the bare object and subsequently plotted against varying values of $\Omega_2$ (in units of $2\pi$ rad) and $\gamma$. This result is shown in Fig.~\ref{fig:fig_scs_2d_cloak}(a) in logarithmic scale. The regions colored with dark blue correspond to significant scattering reduction (i.e. $\sigma^\textrm{scat}_c/\sigma^\textrm{scat}_b\ll 1$), whereas red regions correspond to enhanced scattering from the core-shell geometry. it can be seen that two distinct regions of cloaking can be distinguished, i.e. for $\gamma>0.6$ and for $\gamma\approx 0.3$, and both with values of $\Omega_2/(2\pi)$ between -30 Hz and -10 Hz. In particular, a minimum of -30 dB of $\sigma^\textrm{scat}_c/\sigma^\textrm{scat}_b\ll 1$ is seen around values of $\Omega_2/(2\pi)=-20$ Hz and $\gamma=0.65$.

To isolate the effect of $\Omega_2$ and $\gamma$ on the scattering reduction mechanism, a plot of $\sigma^\textrm{scat}_c/\sigma^\textrm{scat}_b\ll 1$ is given versus $\Omega_2/(2\pi)$ for different values of $\gamma$ in Fig.~\ref{fig:fig_scs_2d_cloak}(c). One can see that one single scattering dip exists for some values of $\gamma$. For instance for $\gamma=0.45;\,0.99$, no cloaking is possible. The minimum cloaking is for $\Omega_2=-20$ Hz and $\gamma=0.65$. Next, $\sigma^\textrm{scat}_c/\sigma^\textrm{scat}_b\ll 1$ is plotted versus $\gamma$ and for different values of $\Omega_2/(2\pi)$. One can see that here two cloaking regimes take place. First, for small values of $\gamma$ around 0.3, a small reduction of the range of -16 dB can be observed for an extended range of $\Omega_2/(2\pi)$. Then, for higher values of $\gamma$, i.e. $\gamma\geq 0.6$, an efficient scattering reduction regime takes place (more than -20 dB). This second cloaking dip is more sensitive to changes of $\Omega_2$, in comparison to the first one, where a redshift can be clearly observed.

To better illustrate the efficiency of the proposed cloak, the far-field scattering patterns (i.e. $|f(\theta)|$) in polar coordinates is shown in Fig.~\ref{fig:fig_scs_2d_cloak}(b), for two specific parameters of $\Omega_2$ and $\gamma$, depicted in Fig.~\ref{fig:fig_scs_2d_cloak}(d) with circles. This plot demonstrates that the spinning fluid is undetectable for all angles [Fig.~\ref{fig:fig_scs_2d_cloak}(c) gives ormalized $|f_c(\theta)/f_b(\theta)|$, with subscripts c and b, denoting, as before, the cloaked and bare object, respectively]. It can be seen also that the high $\gamma$ regime (solid curve) gives better angular SCS reduction than the $\gamma\approx 0.3$ regime, and confirms that an a clockwise rotating shell of small radius can cloak an anti-clockwise spinning object.

\begin{figure}
    \centering
    \includegraphics[width=0.9\columnwidth]{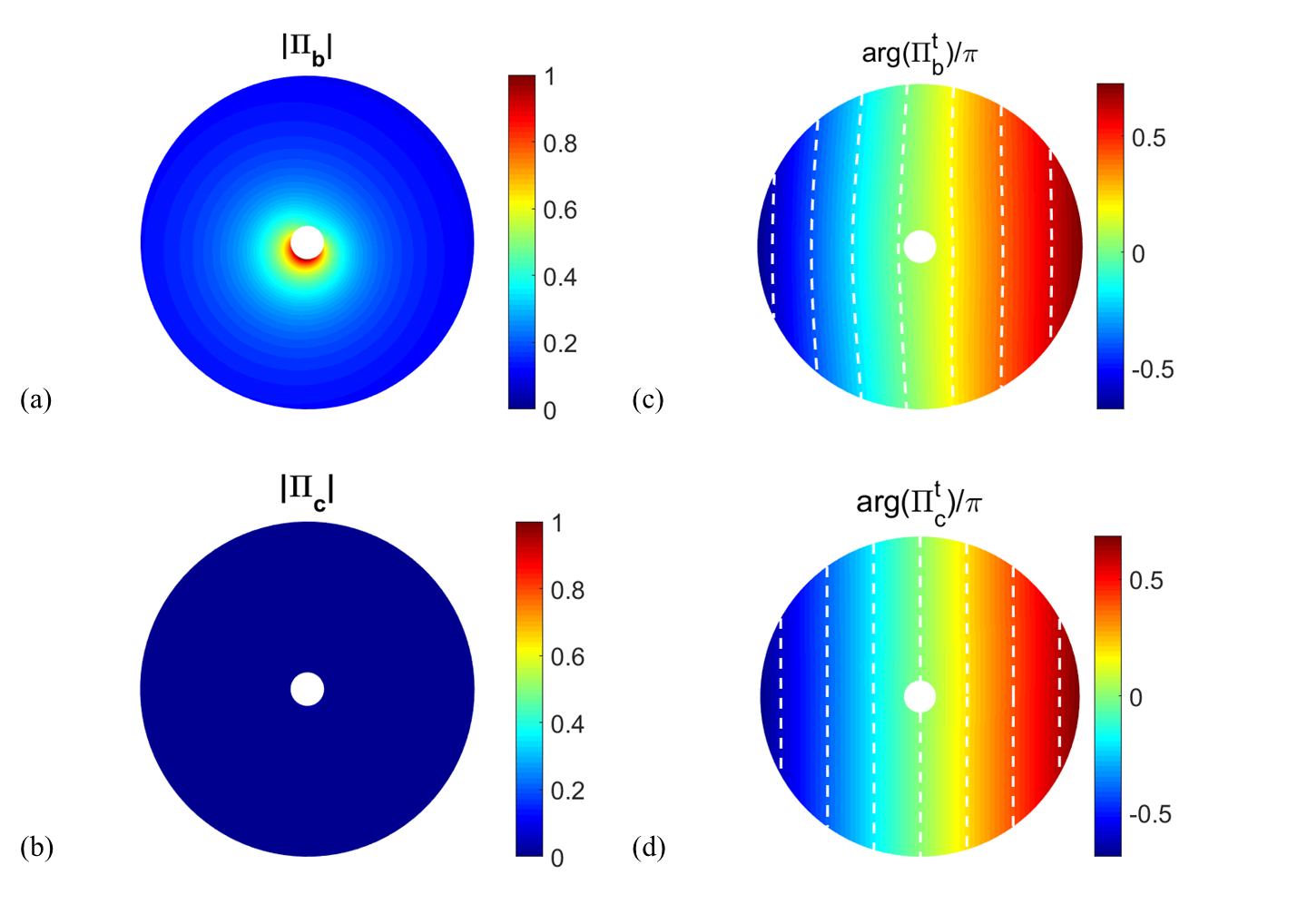}
    \caption{(a) Near-field plot (in arbitrary units) of the acoustic scattered Poynting vector ${\bf \Pi}_b$ (proportional to $|p^\textrm{scat}|^2$) of the bare object of radius $r_1=1$ m, spinning with speed $\Omega_1/(2\pi)=15$ Hz at the frequency $\omega/(2\pi)=16.75$ Hz. (b) Same as in (a) for the cloaked object (${\bf \Pi}_c$), with the shell of radius $r_2=r_1/\gamma$ and $\gamma=0.65$ and spinning frequency $\Omega_2/(2\pi)=-20$ Hz. The physical parameters of the object and shell are equal to those of the surrounding, i.e., water. The phases (normalized with $\pi$) of the total Poynting vectors are given in (c) and (d) for the bare and cloaked object with same parameters as in (a) and (b), respectively. The white dashed lines represent the contours of the phases.}
    \label{fig:fig_nearfield}
\end{figure}

Last, Figs.~\ref{fig:fig_nearfield}(a)-(b) plot the near-field scattered pressure field amplitude (or more precisely the amplitude of the normalized scattered acoustic Poynting vector \cite{tang2018dynamically} , i.e. $|{\bf \Pi}_b|$ and $/|{\bf \Pi}_c|$) in the environment region (region 0) for the bare object and cloaked object, respectively. One can see that a drastic reduction of the scattered fields (about two orders of magnitude), takes place, in the case of an object with spinning coating. The phases (normalized with $\pi$) of the total Poynting vectors are given in Fig.~\ref{fig:fig_nearfield}(c) and Fig.~\ref{fig:fig_nearfield}(d) for the bare and cloaked object with same parameters as in Fig.~\ref{fig:fig_nearfield}(a) and Fig.~\ref{fig:fig_nearfield}(b), respectively. These plots show that the phase of the fields is not distorted in the case of cloaked scenario (straight contour lines, marked with the dashed white lines) whereas for the bare case it is slightly distorted (contour lines are curved due to the enhanced scattering from the spinning object).

\subsection{The Case of a Hard-Wall Object}
\label{sec:appendixb}
Let us first derive the equivalence of the scattering from a hard-wall object and an infinite acoustic impedance object. The hard-wall (rigid) boundary condition for $r=r_1$, namely ${\bf n}\cdot{\bf v}=0$ (or in terms of pressure $1/\rho\,\partial_rp=0$.) The incident field is as usual a plane-wave expressed as in Eq.~(\ref{eq:eq5}) and the scattered pressure is given as in Eq.~(\ref{eq:eq6}). By application of the hard-wall boundary condition at $r_1$, one obtains the expression of the coefficients
\begin{equation}
s_n^{(r)}=\frac{-J'_n\left(\beta_0r_1\right)}{H_n^{(1)'}\left(\beta_0r_1\right)}\, , \,\forall n \in \mathbb{Z}\, .
\label{eq:eqb1}    
\end{equation}
On the other hand, an object of same radius $r_1$, density $\rho_1$, and bulk modulus $\kappa_1$ embedded in a homogeneous medium of density $\rho_0$ and bulk modulus $\kappa_0$, possesses scattering given by
\begin{widetext}
\begin{equation}
s_n=
\begin{vmatrix}
  J_n\left(\beta_{1}r_1\right) & J_n\left(\beta_0r_1\right)\\
  J_n'\left(\beta_{1}r_1\right) & \chi J_n'\left(\beta_{0}r_1\right)
  \end{vmatrix}
   {\begin{vmatrix}
  J_n\left(\beta_{1}r_1\right) & H_n^{(1)}\left(\beta_0r_1\right)\\
  J_n'\left(\beta_{1}r_1\right) & \chi H_n^{(1)'}\left(\beta_{0}r_1\right)
  \end{vmatrix}}^{-1}\; ,
\label{eq:eqb2}
\end{equation}
\end{widetext}
where $\chi=Z_1/Z_0$, and $Z_{0,1}=\sqrt{\rho_{0,1}\kappa_{0,1}}$ the impedance of the object and free-space, respectively.

\begin{figure}[b!]
    \centering
    \includegraphics[width=0.9\columnwidth]{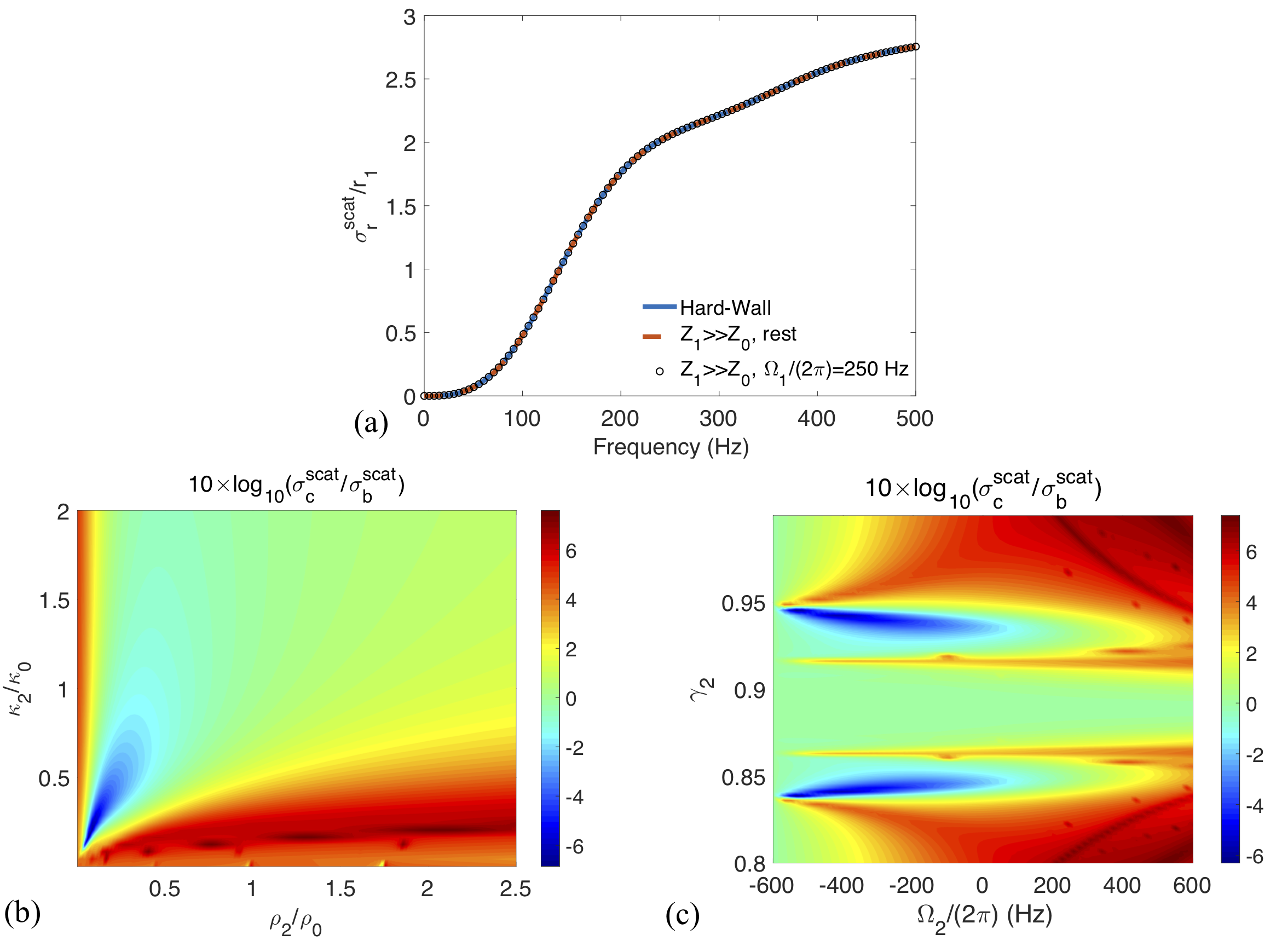}
    \caption{(a) SCS for the hard-wall boundary (blue line), infinite acoustic impedance approximation, i.e., $\sqrt{\rho_1\kappa_1}\rightarrow\infty$ (red dashed line), and spinning infinite acoustic impedance approximation (circles). (b) Cloaking scenario for the hard-wall object of radius $r_1=1$ when using classical SCT scheme, i.e., by varying the density and bulk modulus of the shell of radius $r_2=1.2$ at frequency $\omega/(2\pi)=360$ Hz. (c) Same as in (b) but using a spinning shell of density and bulk modulus equal those of free-space.}
    \label{fig:fig_scs_rigid}
\end{figure}

Now in order to establish the analogy between the hard-wall boundary and the inhomogeneous medium, one must equalize Eqs.~(\ref{eq:eqb1})-(\ref{eq:eqb2}). This can be obtained for all scattering orders, if one ensures that $\chi\rightarrow\infty$, i.e. by assuming an infinite impedance of the object. This is is somehow coherent as the higher impedance leads to enhanced reflection and in this limit the fields cannot penetrate the object, which is an equivalent to hard-wall boundary. This fact is demonstrated in Fig.~\ref{fig:fig_scs_rigid}(a), where the plot of the SCS versus a broadband of frequencies is depicted. It can be also seen from Fig.~\ref{fig:fig_scs_rigid}(a) that rotating a hard-wall object does not change its scattering response, unlike for the case of an acoustic medium with finite impedance. This is mainly due to the fact that there is no flow inside the object (pressure and velocity are zero for $r\leq r_1$) and hence rotating the object does not induce any extra scattering features.

Last, to verify the versatility and robustness of this new kind of SCT-based cloaking, one investigates the possibility to cloak a rigid (hard-wall like) cylindrical object by using only a rotating shell, of same physical parameters ($\rho_2$ and $\kappa_2$) as those of the surrounding medium (water, here). The rigid body can mimic for example a submarine, or any under-water solid (one ignores shear waves here, as only compression waves are investigated). One first coats the rigid object of radius $r_1=1$ with a shell of radius $r_2=1.2$ and one sweeps the density and bulk modulus of the shell, as usually done in SCT cloaking. The normalized SCS is plotted as before, at frequency $\omega/(2\pi)=360$ Hz, and the result is depicted in Fig.~\ref{fig:fig_scs_rigid}(b). On the other hand, one considers to coat the same object with a shell of radius $r_2=r_1/\gamma$ and spinning angular frequency $\Omega_{2}$. In this scenario $\rho_2=\rho_0$ and $\kappa_2=\kappa_0$. So the SCT is induced here purely by spinning effect. The result is depicted in Fig.~\ref{fig:fig_scs_rigid}(c) and it can be clearly seen that comparable scattering cancellation is possible to achieve. The advantage is here that one does not need near-zero or negative density and/or bulk modulus for the cloaking operation [as can be seen from Fig.~\ref{fig:fig_scs_rigid}(b)]. By pure rotation of homogeneous shells, cloaking is made possible. Note one can further improve this scattering rotation with spinning by allowing some freedom for the density and/or bulk modulus of the shell.




\section{concluding remarks}

In summary, a detailed analysis of spinning acoustic objects and their scattering properties is proposed, and acoustic cloaks based on the scattering cancellation technique are designed. Here the main challenge is that the object to conceal (cloak) is not at rest, and experiences rotation along its $z$-axis (for cylinders) at constant angular speed (with a few rotation cycles per second). Scattering by such acoustic rotating objects is physically different from objects at rest, and possesses resonant Mie features at specific frequencies. The cloaking mechanism introduced here presents several advantages in comparison with zero-velocity cloaking, as it may be more useful in realistic applications (where objects are most of the time moving). Using a homogeneous layer of same properties as free-space with a rotation (in the opposite direction to the object) one is able to significantly reduce the scattering from objects with various spinning speeds. It is also shown that using purely a spinning shell, it is possible to cancel the scattering from a rigid (hard-wall) object in a similar manner as optimizing its density and bulk modulus, which shows the versatility of this cloaking mechanism.

Experimental realization of this concept may be within reach readily, as it only requires rotating objects and shells, allowing for interesting applications in scenarios in which it is desirable to suppress the scattering from obstacles that are in a spinning movement (e.g. rotating components of cars or helicopter rotor blades) for noise reduction. The same concept can also be generalized to other classes of waves, such as linear surface water waves, flexural waves in thin-plates or beams.

\section*{Acknowledgements}
The research reported in this manuscript was supported by Baseline Research Fund BAS/1/1626-01-01.

\appendix

\section{Derivation of The Acoustic Equation in Spinning Media}
\label{sec:appendix}

\begin{figure}[t!]
    \centering
    \includegraphics[width=0.6\columnwidth]{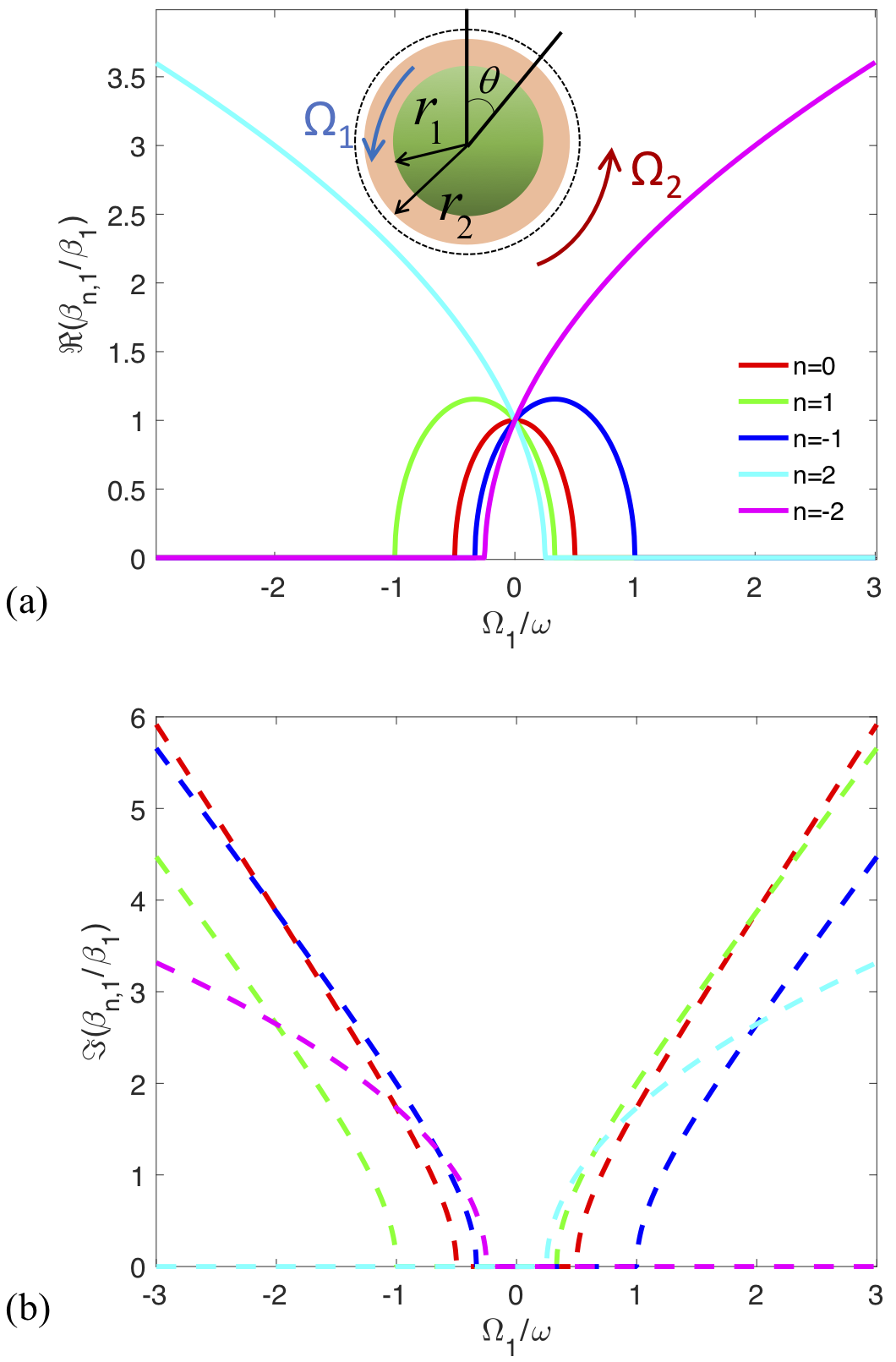}
    \caption{(a) Real  and (b) imaginary parts of the spinning wavenumbers for different orders $n$ versus the rotation coefficient $\alpha_1=\Omega_1/\omega$. The inset of (a) plots the scheme of the multiple layers and the interfaces of an acoustic structure, as well as the rotation directions.}
    \label{fig:fig_scheme}
\end{figure}

Let us consider a uniformly rotating medium, as schematized in Fig.~\ref{fig:fig_scheme}(a). The usual equations of motion (momentum conservation) and continuity (conservation of mass) need be modified \cite{morse1968ku,graham1969effect,censor1971scattering,schoenberg1973elastic,censor1973two,farhadi2018acoustic}. If one considers no shear stresses and body forces, it is shown that the momentum conservation can be written as,
\begin{equation}
    \rho\left(\frac{D{\bf u}}{Dt'}\right)=-{\bf \nabla}' P\; ,
\label{eq:eqa1}
\end{equation}
where the operators $D/Dt'$ and ${\bf \nabla}'$ represent the total time-derivative and the spatial derivative, respectively, in the reference frame $\mathcal{R}'$ associated with the spinning disc. The density $\rho$ is assumed to be constant with respect to time, due to low compressibility and reasonable rates of rotation, as well as small-amplitude sound waves, as usually assumed. Here the total pressure $P$ accounts for the pressure due to acoustic waves as well as to the rotation of the structure and ${\bf u}$ is the total velocity.

In the reference frame $\mathcal{R}$ associated with the laboratory, Eq.~(\ref{eq:eqa1}) is transformed into
\begin{equation}
    \rho\left[\frac{\partial}{\partial t}+\left({\bf u}\cdot{\bf \nabla}\right)\right]{\bf u}=-{\bf \nabla} P\; ,
\label{eq:eqa2}
\end{equation}
with ${\bf u}={\bf u}_0+{\bf v}$, with ${\bf v}$ the velocity of the acoustic wave, and ${\bf u}_0={\bf u}_0({\bf r})$ the bulk velocity, that corresponds to rotation. For a uniform spinning, one has ${\bf u}_0=\Omega r{\bf e}_\theta$, with ${\bf e}_\theta$ the azimutal unit-vector and $\Omega$ the angular velocity. By denoting $P=p_0+p$, with $p_0$ the time-independent pressure due to the frame motion and $p$ the pressure of the acoustic waves (due to the acoustic perturbation). Equation~(\ref{eq:eqa2}) can be expressed as \cite{morse1968ku}
\begin{equation}
    \left[\frac{\partial}{\partial t}+\left({\bf u}_0\cdot{\bf \nabla}\right)\right]{\bf v}+\left({\bf v}\cdot{\bf \nabla}\right)\cdot{\bf u}_0=-\rho^{-1}{\bf \nabla} p\; .
\label{eq:eqa3}
\end{equation}
For the mass conservation equation, a similar reasoning permits to show that it can be expressed in the laboratory frame $\mathcal{R}$ as
\begin{equation}
    \left[\frac{\partial}{\partial t}+\left({\bf u}_0\cdot{\bf \nabla}\right)\right]p+c^2\rho{\bf \nabla}\cdot{\bf v}=0\; ,
\label{eq:eqa4}
\end{equation}
using the fact that ${\bf \nabla}\cdot{\bf u}_0=0$ and noting $c=\sqrt{\kappa/\rho}$, with $\kappa$ the bulk modulus of the structure.

Similarly, the boundary conditions at the interface between two spinning media (or a spinning media and a medium at rest) shall be modified \cite{morse1968ku}. For instance, the pressure $p$ is continuous across the interface. For the second boundary condition, that is the normal component of the velocity (${\bf n}\cdot{\bf v}$), in media at rest, it was shown in Ref. \cite{morse1968ku} that it should be replaced in the moving media by the displacement $\psi$, that is related to the pressure through the modified relation
\begin{equation}
\rho\left(\frac{\partial}{\partial t}+v_{n_1}\frac{\partial}{\partial n_1}\right)\psi_{n_2}=-\frac{\partial p}{\partial n_2}\, ,
\label{eq:eqa5}    
\end{equation}
with $v_{n_1} = {\bf v} \cdot {\bf n}_1$ and $\psi_{n_2} = {\bf \Psi} \cdot {\bf n}_2$, where ${\bf n}_1$ is the direction of the flow velocity and ${\bf n}_2$ is the normal to the considered interface. It is also assumed here that a very thin membrane separates both fluids from mixing, and that the spinning of both fluids is thus independent.
\nocite{*}

\bibliography{apssamp}

\newpage


\newpage

\newpage

\newpage

\newpage

\newpage

\end{document}